\documentclass[prl,twocolumn,showpacs,floatfix,superscriptaddress]
{revtex4}

\usepackage{graphicx}
\usepackage{amsmath}
\usepackage{amsfonts}

\begin{document}

\title{Dissipation: The phase-space perspective}

\author{R. Kawai}
\affiliation{Department of Physics, University of Alabama at Birmingham,
Birmingham, AL 35294}

\author{J. M. R. Parrondo}
\affiliation{Departamento de F{\'i}sica At{\'o}mica, Molecular y
Nuclear and
GISC, Universidad Complutense de Madrid, 28040-Madrid, Spain}

\author{C. Van den Broeck}
\affiliation{University of Hasselt, B-3590 Diepenbeek, Belgium}

\date{\today}

\begin{abstract}

We show, 
through a refinement of the work
theorem,
that the average dissipation, upon perturbing 
a Hamiltonian system arbitrarily far out of
equilibrium in a transition between two
canonical equilibrium states, is
exactly given by $\langle W_{diss} \rangle = \langle W \rangle
-\Delta F =kT D(\rho\|\widetilde{\rho})= kT \langle \ln
(\rho/\widetilde{\rho})\rangle$, where  $\rho$
 and $\widetilde{\rho}$ are the phase space density of the
  system measured at  the same intermediate but otherwise arbitrary
point in time, for the forward and backward process.
$D(\rho\|\widetilde{\rho})$ is the relative entropy of $\rho$ versus
$\widetilde{\rho}$. This result also implies general inequalities, which
are significantly more accurate than the
second law and include, as a special case, the celebrated Landauer
principle on the dissipation involved in irreversible computations.

\end{abstract}


\maketitle

Since the pioneering work of Boltzmann, the search for 
an exact microscopic expression of dissipation has been at the heart of
(nonequilibrium) statistical mechanics. In this letter, we derive
such an expression from first principles, through a 
refinement of
the recently discovered work theorem by Jarzynski and
Crooks~\cite{jarzynski}.

The second law of thermodynamics stipulates that  the average
mechanical work $\langle W \rangle$ needed to move a system  in
contact with a heat bath at temperature $T$, from one equilibrium
state $A$ into another equilibrium state $B$, is at least equal to
the free energy difference between these states: $\langle W \rangle
\geq\Delta F=F_B-F_A$. The equality is reached for a quasi-static
process. The extra work $\langle W \rangle-\Delta F$ is often
referred to as the dissipated work. Contrary to the reversible work,
the dissipated work depends on how the transition between the states
is realized. Typically, one or more external control parameters are
changed in time, following a specific protocol, between initial an
final values. The dissipated work will depend on this protocol,
which can in principle bring the system arbitrarily far out of
equilibrium. Amazingly,  there exists an exact, simple
and  compact microscopic expression for this dissipation. The key is
to consider the protocol  and its time-reversed version. We will
refer with a superscript tilde to all the quantities measured in the
time-reversed protocol. The central result is then the following:
$\langle W _{diss}\rangle =\langle W \rangle -\Delta F
=kTD(\rho||\tilde\rho)= kT \langle \ln
(\rho/\widetilde{\rho})\rangle$, where $\rho$ and $\widetilde{\rho}$
are the phase space densities of the system measured at the same
intermediate but otherwise arbitrary point in time, in the forward and
backward protocol, respectively. $D(\rho\|\widetilde{\rho})$ is the
Kullback-Leibler distance~\cite{cover}, also called relative
entropy, of $\rho$ versus $\widetilde{\rho}$.

The derivation is similar in spirit to that of the Jarzynski and
Crooks equalities \cite{jarzynski}. Consider a Hamiltonian
$H(q,p;\lambda)$ where $(q,p)$ represents the set of position and
momentum variables of the system under consideration and $\lambda$
is a control parameter which is varied from an initial value
$\lambda_A$ to a final value $\lambda_B$ according to a protocol
$\lambda(t)$ controlled by an external agent. The system is initially
assumed to be in canonical equilibrium at
temperature $T$ at the value $\lambda_A$ of the control parameter and,
along
the protocol, is completely isolated, i.e., no energy is exchanged other
than the work $W$ performed by the external agent on the system. In the
time-reversed scenario, the system is initially at canonical
equilibrium at the same temperature $T$, but  at the value
$\lambda_B$ of the control parameter, which is now changed according
to the exact time-reversed protocol.

\begin{figure}[b]
\includegraphics[width=2in]{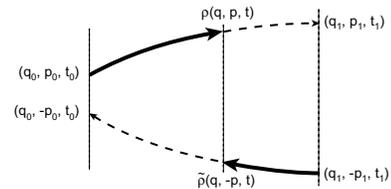}
\caption{Space-time coordinates for the forward and backward
trajectories. } \label{fig:traject}
\end{figure}

We first  set out to calculate the work $W(q,p\,;t)$ done along the
whole process, for the specific phase trajectory that passes through
the  phase point $(q,p)$  at time $t$. Since the dynamics are
deterministic, there is precisely one such trajectory. Let us call
$(q_0,p_0)$ and $(q_1,p_1)$ the corresponding initial and final
phase points.  Note also that there is a one-to-one correspondence with
the time-reversed trajectory in the time-reversed protocol which,
starting from $(q_1,-p_1)$, goes through $(q,-p)$  and finally  into
$(q_0,-p_0)$, cf. Fig.~\ref{fig:traject}.  For simplicity of
notation, we will use the forward time to express times in both
forward and backward scenarios. By conservation of total energy, one
has that:
\begin{equation}
 W(q,p;t) = H(q_1,p_1;\lambda_B)-H(q_0,p_0;\lambda_A)\\
\label{eq:totalWork}
\end{equation}

Now, since the phase
space density is conserved along any Hamiltonian trajectory, one has, in
both forward and backward process:
\begin{eqnarray}
\rho(q,p,t) = \rho(q_0,p_0,t_0)
 &=&\frac{\exp[-\beta H(q_0,p_0;\lambda_A)]}{Z_A}\label{eq:rho}\\
\widetilde{\rho}(q,-p,t) = \widetilde{\rho}(q_1,-p_1,{t}_1) 
&=&\frac{\exp[-\beta
H(q_1,-p_1;\lambda_B)]}{Z_B} \nonumber
\end{eqnarray}
where $Z_A$ and $Z_B$ are partition functions at the equilibrium
states $A$ and $B$, respectively.  These expressions
allow us to
eliminate the Hamiltonian (which is supposed to be even in the
momenta, or more precisely time-reversible) at initial and final times
in favor of the phase space density at any intermediate time point.
 Eq.~(\ref{eq:totalWork}),
yields the following generalized Crooks relation:
\begin{equation}
 \exp\left\{\beta [W(q,p;t)-\Delta F]\right\} =
\frac{\rho(q,p,t)}{\widetilde{\rho}(q,-p,t)}
\label{eq:crooks1}
\end{equation}
where $\Delta F = -kT (\ln Z_B -  \ln Z_A)$ is the free energy
difference between the final and initial equilibrium states. If we
now rewrite Eq. (\ref{eq:crooks1}) as follows:
\begin{equation}
 W(q,p;t) - \Delta F = kT \ln
\frac{\rho(q,p,t)}{\widetilde{\rho}(q,-p,t)}\;, \label{eq::crooks2F}
\end{equation}
the average work reads:
\begin{eqnarray}
\langle W \rangle -\Delta F
&=& kT \int dq dp\,\, \rho(q,p,t) \ln
\frac{\rho(q,p,t)}{\widetilde{\rho}(q,-p,t)} \nonumber\\
&=& kT D\big(\rho(q,p,t)\|\widetilde{\rho}(q,-p,t)\big).
\label{eq:newEquality}
\end{eqnarray}

We conclude that the dissipated work is fully revealed by the phase
space density of forward and backward processes at any intermediate
time of the experiment. It is particularly interesting to note that
this dissipation, cf.~r.h.s.~of Eq.~(\ref{eq:newEquality}), takes
the form of the relative entropy (Kullback-Leibler
distance~\cite{cover}) $D(\rho\|\widetilde{\rho})=\int dq dp\,\,
\rho \ln( \rho/\widetilde{\rho})$ between the forward and backward
probability distributions $\rho$ and $\tilde{\rho}$. This simple
result calls for a number of more specific comments. First, since a
relative entropy is strictly non-negative,  we conclude that the
dissipation cannot be negative, in agreement with the second law.
Second, the dissipation results from the asymmetry between the
forward and backward protocols: it is zero only when $\rho(q,p,t)=
\widetilde{\rho}(q,-p,t)$. In fact, Stein's lemma \cite{cover}
relates $D(\rho\|\widetilde{\rho})$  directly to the difficulty of
statistically distinguishing forward versus backward trajectories.
This is consistent with the general observation that dissipation is
the result of the breaking of detailed balance  \cite{maes}. The
above expression is  also consistent with a  proposal, linking the
time-asymmetry of the Kolmogorov-Sinai entropy to the entropy
production of the dynamical system  \cite{gaspard}. Third,  the
total dissipation, cf.~the l.h.s.~of Eq.~(\ref{eq:newEquality}), is
obviously a constant, independent of time. Yet the densities in the
r.h.s. of Eq. (\ref{eq:newEquality}) can be evaluated at any
intermediate time $t$.
This time-independence  follows from the observation that the relative
entropy of
densities obeying the same Liouville equation, is constant in time
\cite{mackey}.  Fourth, the evaluation of the dissipated work in
general requires  full knowledge of the phase space density, even
though only at one particular instant of time.   That such detailed
information may be needed is consistent with the generality of the
result, which is valid no matter how far the system is driven  out
of equilibrium. However, one can get away from this apparently
stringent requirement, by invoking  the chain rule for relative
entropy \cite{cover}. According to this rule, the relative entropy
decreases upon coarse graining. The equality
Eq.~(\ref{eq:newEquality}) is then replaced by an inequality.  It is
instructive to give a direct derivation of this result.

Consider a partition of the entire phase space, consisting of $K$
non-overlapping subsets $\chi_j,\, (j=1,\dots,K)$. We introduce  the
corresponding coarse grained phase densities
\[
\rho_j = \int_{\chi_j} \rho(q,p) dqdp\,; \quad
\widetilde{\rho}_j = \int_{\widetilde{\chi}_j}
\widetilde{\rho}(q,-p,t) dq dp
\]
 where the $\widetilde{\chi}_j$ is identical to $\chi_j$, apart from the
inversion of
all momenta.  By integration of Eq.~(\ref{eq:crooks1}) over the set
$\chi_j$, we obtain the following detailed Jarzynski equality:
\begin{equation}
 \langle e^{-\beta W} \rangle_j =
\frac{\int_{\chi_j}\rho(q,p,t) e^{-\beta W(q,p;t)} dq dp}{\rho_j} =
\frac{\widetilde{\rho}_j}{\rho_j} e^{-\beta \Delta F}
\label{eq:subjarzynski}
\end{equation}
By Jensen's inequality,  Eq.~(\ref{eq:subjarzynski}) implies a
second-law
like inequality:
\begin{equation}
\langle W \rangle_j \geq \Delta F + k T \ln \left
(\frac{\rho_j}{\widetilde{\rho}_j}\right ) \geq
-\langle \widetilde{W} \rangle_j\, .
\label{eq:subSecondLaw}
\end{equation}
where we have included, for later reference, the inequality that arises
by considering the backward process.
Finally, by performing an average over the different subsets, one
finds:
\begin{subequations}
\label{eq:avgWork}
\begin{eqnarray}
\langle W \rangle &=& \sum_j \rho_j \langle W \rangle_j \geq \Delta F
+kT  D(\rho_j\|\widetilde{\rho}_j)
\label{eq:avgWork1} \\
{\langle \widetilde{W} \rangle} &=& \sum_j \widetilde{\rho}_j
\langle\widetilde{W} \rangle_j \geq -\Delta F
+ kT D(\widetilde{\rho}_j \|\rho_j)\, .
\label{eq:avgWork2}
\end{eqnarray}
\end{subequations}
where the discrete version of
relative entropy is defined by
$D(\rho_j\|\widetilde{\rho}_j)=\sum_j \rho_j
\ln(\rho_j/\widetilde{\rho}_j)$.

We conclude that, when full information of the phase density is not
available, a coarse grained relative entropy still provides a lower
bound for the dissipative work, significantly improving the
classical one given by the second law. How well this bound
approaches the total dissipation will depend on how far the process
is from the quasi-static regime. In particular, in the latter case,
any partition will do, and the relative entropy is always
identically zero. More interestingly, one expects that a
coarse-grained partition will suffice in case of
 separation of time scale between fast and slow variables. Indeed, in
this case and for a protocol on the slow time scale, the fast variables
will be essentially at equilibrium and all the dissipation is  captured
in the
time-asymmetry  contained in the slow variables.
We note however that in this
reduced set of variables, trajectories can, unlike in full phase
space, cross each other. A more detailed analysis reveals that  full
information is then not captured by measurement at a single
 time, but should in general be carried out at all times, in
agreement with, e.g., the entropy production of Markovian processes
\cite{maes,gaspard,lebowitz}.

To illustrate the power and usefulness of our results, we turn to a
number of specific examples. We first consider the quenching of a
system described by the Hamiltonian  $H(x;\lambda) $, in contact
with a heat bath at temperature $T$. The equilibrium probability
distribution to observe the state  $x$ is given by a Boltzmann
distribution $ p(x;\lambda) \equiv \exp ( - \beta H(x;\lambda)
)/Z_\lambda$ with  $Z_\lambda$ the normalization factor (partition
function).  We now perturb this equilibrium by the following
irreversible quench:  the control parameter  is changed
instantaneously from the value $\lambda$ to the value
$\lambda^\prime$  at a specific time and  the experiment terminates
at any later time. In the backward process, we start from
equilibrium at  $\lambda^\prime$ and quench back to $\lambda$. Since
the state $x$ does not change during the instantaneous quench, the
work  in the forward process is:
\begin{equation}
\label{Wquench1}
\langle W \rangle  = \int  \textrm{d}x \;\;[H(x;\lambda^\prime)-
H(x;\lambda)] \;\;p(x;\lambda)
\end{equation}
Now consider a partition, infinitely fine in the position coordinates
(disregarding all
the other degrees of freedom, in particular those of the heat
bath) and measure the coarse grained distribution $p(x,t)$ at the time
of
quench.
We note that the distributions prior to the quench  are the
equilibrium distribution from which one started, i.e.,
$p(x;\lambda)$  and $p(x;\lambda^\prime)$ for forward and
backward scenario respectively. Turning to Eq.
(\ref{eq:subSecondLaw}), the role of $\rho_{j}$ and
$\widetilde{\rho}_{j}$
are thus played by  $p(x;\lambda)$ and
$p(x;\lambda^\prime)$, hence:
\begin{equation}
\langle W \rangle\ge \Delta F + kT\int \textrm{d}x\;\;
p(x;\lambda)\;\;\ln \frac{p(x;\lambda)}{p(x;\lambda')}
\label{eq:Wquench2}
\end{equation}
Using the Boltzmann probability distributions together with $\Delta
F=-k T \ln (Z_{\lambda^\prime}/Z_\lambda)$,  a comparison with
Eq.~(\ref{Wquench1}) reveals that  the equality sign holds in
Eq.~(\ref{eq:Wquench2}). We conclude that in this  case the
coarse-grained partition captures the full dissipation.

\begin{figure}[b]
\includegraphics[width=3in]{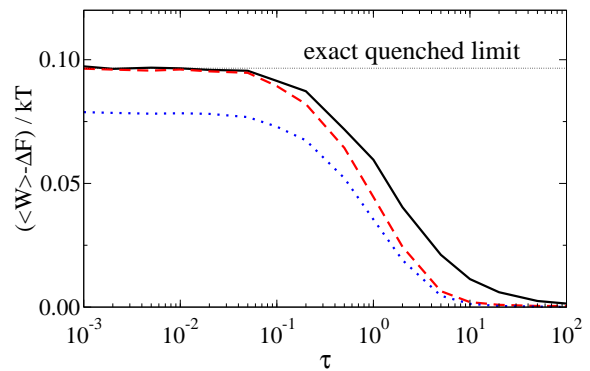}
\caption{Dissipation of Brownian particles in a harmonic potential
with spring constant varying from $\kappa=2$ to $\kappa^\prime=1$
during a time interval $\tau$.  Solid line is dissipative work
directly obtained from simulation.  Dashed and dotted lines indicate
the lower bound of the dissipative work estimated from the relative
entropy.} \label{fig:Wdis}
\end{figure}

Turning to a more complicated situation, we consider an overdamped
Brownian particle in contact with a heat bath at temperature $T$,
moving in a harmonic potential whose spring constant varies from
$\kappa$ to $\kappa^\prime$ during a finite time $\tau$. For $\tau
\rightarrow 0$, one recovers the quenching experiment described
above with $\langle W \rangle$ given by Eq.~(\ref{Wquench1}) (with a
harmonic Hamiltonian). For $\tau \rightarrow \infty$, one approaches
the quasi-static limit with $\langle W \rangle=\Delta F$. In
Fig.~\ref{fig:Wdis} we compare the dissipative work $\langle W \rangle -
\Delta F$, obtained from Langevin simulations, with the
relative entropy measured at the middle of the transition with a
fine partition ($\Delta x=0.1$) and a coarse partition ($\Delta
x=1.0$).  The relative entropy is always below the dissipative work,
consistent with Eq.~(\ref{eq:avgWork1}). For the fine partition the
relative entropy coincides with the dissipative work as $\tau$
approaches the quenched limit, in agreement with
Eq.~(\ref{eq:Wquench2}). Note that the refinement of the partition
in estimating the dissipation  is most effective close to the
quenched limit.

\begin{figure}[b]
\begin{center}
\centerline{\includegraphics[width=2.7in]{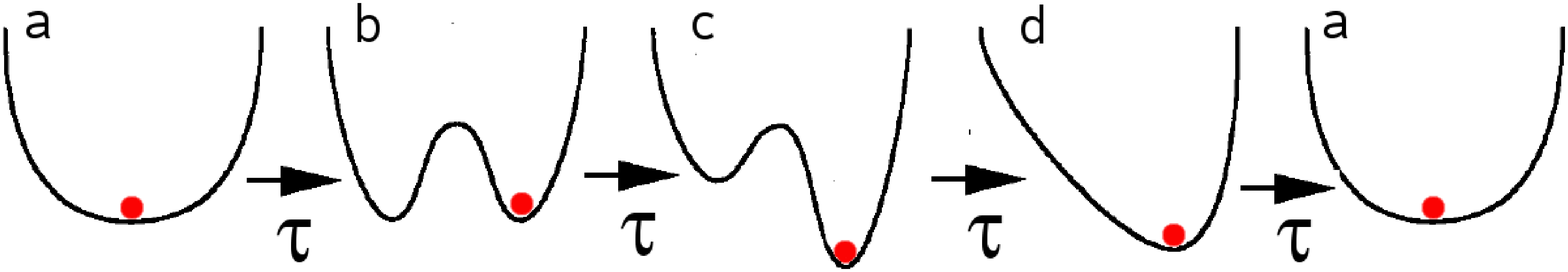}}
\vspace{0.05in}

\centerline{\includegraphics[width=3.0in]{fig3b.eps}}
\caption{
{\em Top:} Cyclic variation of a model potential.
 The arrows indicate
the
forward protocol corresponding to the Szilard engine.  The backward
process is a model for recording and erasing information.
{\em Bottom:} Partitioned average work as a function of processing time
$\tau$: $\langle W \rangle_R$ (solid line) and $-\langle \widetilde{W}
\rangle_R$ (dashed line).  The bound
(\ref{eq:subSecondLaw}) using the coarse grained distribution is
shown by the dotted line.}
\label{fig:pw}
\end{center}
\end{figure}

In our final illustration, we show that Eqs.~(\ref{eq:avgWork})
include as a special case the celebrated Landauer principle on the
minimal dissipation of irreversible computations. We consider a
Brownian computer~\cite{landauer,bennett,others}, consisting of a
one-dimensional overdamped Brownian particle at temperature $T$ in a
time-dependent potential varied by an external agent according to a
given cyclic protocol shown in Fig. \ref{fig:pw}. Since it involves
spontaneous symmetry breaking followed by forced symmetry breaking,
this process is analogous to the Szilard engine~\cite{parrondo},
whereas the reverse, starting from {\bf b}, is analogous to the
Landauer's restore-to-zero process \cite{leff,landauer,bennett}.

The coarse resolution measurement is made at the stage {\bf b} of the
forward cycle by partitioning position space into two sets,
$\chi_R=\{x:x \geq 0\}$ and $\chi_L=\{x:x < 0\}$  (See Fig.
\ref{fig:pw}). In the forward process, we have by symmetry that
$p_R=p_L=1/2$. In the backward process the large majority of
trajectories  will be forced by the external bias towards the
location of the cell $\chi_R$ at stage {\bf d}. However, since
the height of the barrier is finite, trajectories can still
thermally cross over to $\chi_L$ before reaching the filtering
stage {\bf b}.  Therefore, the probabilities $\widetilde{p}_R$ and
$\widetilde{p}_L$, while being close to $1$ and $0$ for strong
forcing,
will otherwise  depend on the applied force, barrier height,
temperature and processing speed.

In this example, we focus on the validity of
Eq.~(\ref{eq:subSecondLaw}), i.e., the average work for each
macroscopic trajectory, $R$ or $L$. For strong forcing,
$\widetilde{p}_R=1$, Eq.~(\ref{eq:subSecondLaw}) reads:
\begin{equation} \langle W\rangle_R \ge -kT\ln 2 \ge
-\langle\widetilde W\rangle_R\;\;\;\;(\widetilde{p}_R=1).
\label{eq:szilard}
\end{equation}
This expression includes Landauer's principle in the case of the
backward process, namely, that the erasure of one bit of information
must be accompanied by the dissipation of an energy $kT\ln 2$.
 For the Szilard engine, i.e., the forward
process, one recovers the apparent violation of the second law by
$R$ trajectories 
when the equal sign in Eq.~(\ref{eq:szilard})
holds: $\langle W\rangle_R = -kT\ln 2$. Concomitantly, along the $L$
path, we have $\tilde{p}_L \rightarrow 0$, or, more precisely,
$\tilde{p}_L\approx \exp(-V/kT)$, $V$ being the height of the
barrier, and the lower bound given by Eq.~(\ref{eq:avgWork1}) is,
approximately, $\langle {W}\rangle_L>V$. The $L$  trajectories
correspond to a ``wrong" measurement in the Szilard
engine~\cite{parrondo}, and they dissipate an energy significantly
bigger than the energy $kT\ln 2$ extracted from the thermal bath in
the $R$ trajectories. Consequently, the overall dissipation for the
engine is positive, in accordance with the second law.

In Fig. \ref{fig:pw} we show the results of numerical simulations of
the corresponding overdamped Langevin equation.  Average work
performed by the Brownian particles residing in the right well at
the stage \textbf{b} in the forward and backward processes is
plotted for different values of the processing time $\tau$.  The
bound using $p_R$ and $\widetilde{p}_R$ measured directly in the
simulation is also plotted. The bound is always between $\langle W
\rangle_R$ and $-\langle \widetilde{W} \rangle_R$ in agreement with
Eq.~(\ref{eq:subSecondLaw}). Note that the Landauer principle,
contained in  (\ref{eq:szilard}), breaks down in the quasi-static
limit $\tau \rightarrow \infty$,  since the Brownian particle
equilibrates by crossing the barrier. In this process, the stored
information is lost, $\widetilde{p}_R \rightarrow 1/2$, and the
(dissipated) work goes to zero. Nevertheless, the inequality
(\ref{eq:subSecondLaw}) is always satisfied.

As the Jarzynksi equality itself has generated a lot of debate,  a
critical discussion of the above theory is in place. The term
dissipation is usually associated to entropy production. To make
this connection, we note that the system is not at equilibrium at
the final stage of the (forward) experiment. We can however
reconnect it to a heat reservoir and let it relax to its canonical
equilibrium state for $\lambda=\lambda_B$ and  temperature $T$. In
doing so, it will exchange the dissipated work under the form of
heat with the bath, resulting in an entropy production $\Delta S=
k \langle \ln (\rho/\widetilde{\rho})\rangle$. One could ask whether
the disconnection or reconnection of the system with the bath adds
significant terms to the work and/or the free energy.  Apart from
the answers given to these issues in the context of the Jarzynski
equality itself \cite{jarzynksireply}, we are concerned here with
the average total work, which is much larger than these energies for
large systems and long enough operation times. Finally, the above
derivation relies on a continuous transformation of the Hamiltonian,
excluding, for example, free expansion. This limitation reflects the
need for  considering the time-reversed process.

This collaboration was supported by the StochDyn program of the European
Science Foundation.
J.M.R.P. acknowledges financial support from Ministerio de
Educaci\'on y Ciencia (Spain), grant FIS04-271, and from BCSH, grant
UCM PR27/05-13923- BSCH.

\end{document}